\documentclass[journal=jacsat,manuscript=article]{achemso}

\usepackage[version=3]{mhchem} 
\usepackage{algorithm} 
\usepackage{algpseudocode} 
\usepackage{subcaption}
\usepackage{braket}
\SectionNumbersOn  

\usepackage{soul}
\usepackage{xcolor}
\usepackage{soul}
\usepackage{hyperref}

\usepackage{xr}
\makeatletter
\newcommand*{\addFileDependency}[1]{
  \typeout{(#1)}
  \@addtofilelist{#1}
  \IfFileExists{#1}{}{\typeout{No file #1.}}
}
\makeatother

\newcommand*{\myexternaldocument}[1]{%
    \externaldocument{#1}%
    \addFileDependency{#1.tex}%
    \addFileDependency{#1.aux}%
}
\usepackage{xr-hyper}
\myexternaldocument{supp_info} 

\author{Nonia Vaquero-Sabater}
\affiliation{Donostia International Physics Center (DIPC), 20018 Donostia, Euskadi, Spain}
\alsoaffiliation{Polimero eta Material Aurreratuak: Fisika, Kimika eta Teknologia Saila, Kimika Fakultatea, Euskal Herriko Unibertsitatea (EHU), PK 1072, 20080 Donostia, Euskadi, Spain}
\author{Abel Carreras}
\affiliation{Donostia International Physics Center (DIPC), 20018 Donostia, Euskadi, Spain}
\author{Lukas Broers}
\affiliation{RIKEN Center for Computational Science (R-CCS), Kobe 650-0047, Japan}
\author{Tomonori Shirakawa}
\affiliation{RIKEN Center for Computational Science (R-CCS), Kobe 650-0047, Japan}
\alsoaffiliation{RIKEN Center for Quantum Computing (RQC), Wako 351-0198, Japan}
\alsoaffiliation{RIKEN Center for Interdisciplinary Theoretical and Mathematical Sciences (iTHEMS), Wako 351-0198, Japan}
\alsoaffiliation{RIKEN Pioneering Research Institute (PRI), Wako 351-0198, Japan}
\author{Seiji Yunoki}
\affiliation{RIKEN Center for Computational Science (R-CCS), Kobe 650-0047, Japan}
\alsoaffiliation{RIKEN Center for Quantum Computing (RQC), Wako 351-0198, Japan}
\alsoaffiliation{RIKEN Cluster for Pioneering Research (CPR), Wako 351-0198, Japan}
\alsoaffiliation{RIKEN Center for Emergent Matter Science (CEMS), Wako 351-0198, Japan}
\author{David Casanova}
\affiliation{Donostia International Physics Center (DIPC), 20018 Donostia, Euskadi, Spain}
\alsoaffiliation{IKERBASQUE, Basque Foundation for Science, 48009 Bilbao, Euskadi, Spain}
\email{david.casanova@dipc.org}

\title[]{Noise and Configuration Recovery Impact on Quantum Selected Configuration Interaction}

\abbreviations{}

\begin{document}


\begin{tocentry}
\end{tocentry}

\begin{abstract}
Quantum-selected configuration interaction (QSCI) is a promising hybrid quantum-classical approach in which a quantum device generates configurations for subsequent classical diagonalization.
Here, we analyze the performance of QSCI combined with the local unitary cluster Jastrow (LUCJ) ansatz, focusing on the interplay between ansatz expressivity, sampling, noise, and configuration recovery.
Using the dissociation of \ce{N2} in a large active space as a benchmark, we show that noiseless LUCJ sampling produces compact and biased configurational spaces, limiting the accuracy of the resulting CI energies, particularly in strongly correlated regimes.
By introducing a simple noise model, we demonstrate that sampling noise can enhance Hilbert-space exploration by generating additional configurations beyond those supported by the ideal ansatz.
When combined with configuration recovery, this leads to systematically improved energies.
Moreover, recovery alone (starting from randomly generated configurations) can efficiently construct accurate CI spaces, highlighting its central role in QSCI.
\end{abstract}

\clearpage

\section{Introduction}
Solving the time-independent Schr\"odinger equation for the molecular Hamiltonian has long been a central objective of theoretical quantum chemistry, particularly in the pursuit of accurate electronic ground-state energies.
The fundamental difficulty of this problem stems from the exponential growth of the Hilbert space with the number of electrons, which renders an exact treatment intractable for all but the smallest systems.
Over the decades, a wide range of approximate electronic structure methods has been developed to overcome this limitation, typically by sacrificing some degree of exactness in exchange for improved computational scalability.
Most correlated approaches are constructed on top of a mean-field reference, the Hartree-Fock (HF) solution, which provides a single-determinant approximation to the many-electron wavefunction.
From this starting point, electron correlation can be incorporated through systematic expansions or corrections,\cite{Helgaker:2000,Cramer:2002} such as coupled cluster (CC) theory, configuration interaction (CI) methods, and many-body perturbation theory, notably M{\o}ller–Plesset perturbation theory.

In this context, quantum computers represent a promising approach to mitigating the system-size problem in quantum chemistry, as they naturally operate within an exponentially large Hilbert space. 
As a result, quantum chemistry has become a promising application area for quantum computing.\cite{McArdle:quantum-chem:2020,Cao:quantum-chem:2019}
At present, quantum computing is in the noisy intermediate-scale quantum (NISQ) era,\cite{Preskill2018quantumcomputingi} which requires practical quantum algorithms to exhibit a certain degree of noise resilience.
This requirement has motivated the development of hybrid quantum-classical approaches.\cite{bharti2022}
Variational algorithms have emerged as a key area of research in recent years, driving much of the work in quantum computing applied to quantum chemistry.\cite{Peruzzo:2014,McClean_2016,Grimsley2019}
However, the application of these algorithms to realistic quantum chemistry problems remains challenging, as the required circuit depths are often prohibitively large,\cite{Hastings2015} and noise in current quantum hardware severely limits the size of the systems that can be practically implemented.\cite{carreras2025limitationsquantumhardwaremolecular}

These limitations highlight the need for quantum algorithms that can efficiently represent correlated wavefunctions while maintaining shallow circuit constructions.
A natural framework for discussing wavefunction representations in this context is CI theory.
In CI, electronic states are expressed as a linear combination of Slater determinants (or configuration state functions),
\begin{equation}\label{eq:ci_wv}
|\Psi\rangle = \sum_{i\in S} c_i |\phi_i\rangle,
\end{equation}
where the coefficient $c_i$ quantifies the contribution of configuration $|\phi_i\rangle$ to the target state $|\Psi\rangle$, and $S$ denotes the set of configurations included in the expansion.
In the limiting case where $S$ comprises all possible configurations within a given one-particle basis set, the expansion becomes equivalent to the FCI solution.
However, the factorial scaling of FCI with system size renders it computationally prohibitive for all but the smallest systems, necessitating truncations of the configurational space $S$.
To date, the largest FCI calculation reported involved approximately 1.3 trillion determinants and relied on a highly parallelized computational implementation.\cite{Gao_2024}
A common strategy to reduce the size of the CI expansion is to restrict configurations according to their excitation degree relative to a reference (or multireference) determinant, for instance by including only single excitations (CIS), singles and doubles (CISD), and so forth.
Alternatively, CI wavefunctions can be constructed by selecting configurations deemed important for describing the target states, independently of their excitation level.
Following this philosophy, a wide variety of selected CI approaches have been developed, many of which are capable of recovering a substantial fraction of the correlation energy at a significantly reduced computational cost.\cite{Szabo1996,Holmes2016b,Holmes2016_HCI,Sharma2017}

Despite the success of selected CI approaches, the computational cost associated with the identification and ranking of important configurations grows rapidly with system size, ultimately becoming prohibitive for medium- to large-scale problems.
The bottleneck lies in the combinatorial nature of the configurational search, which demands increasingly sophisticated heuristics and screening criteria to remain tractable.

To address this issue, Quantum-Selected Configuration Interaction (QSCI)\cite{Kanno2023,Alexeev2024} was recently introduced.
In QSCI, the configurational search is delegated to quantum hardware by sampling a parameterized ansatz state.
In principle, the quantum device can explore the Hilbert space more efficiently than classical algorithms, potentially outperforming conventional selection strategies.

A central bottleneck of this strategy is the so-called quantum sampling problem,\cite{Reinholdt2025} which originates from the probabilistic nature of measurements on the prepared quantum state. As the algorithm iteratively refines the CI subspace, the probability distribution becomes increasingly concentrated on a subset of already-identified, high-weight determinants. Consequently, repeated measurements predominantly reproduce the same configurations, while the discovery of new, low-probability yet potentially important determinants requires a disproportionately large number of shots. This imbalance leads to a rapidly increasing sampling cost and limits the efficiency with which the configurational space can be systematically expanded.
Despite this challenge, QSCI has demonstrated remarkable performance in nontrivial molecular systems, including the computation of the triple-bond dissociation profile of \ce{N2} and the ground states of [2Fe-2S] and [4Fe-4S] clusters.\cite{RobledoMoreno2025} Furthermore, recent refinements of the overall workflow, such as improved state preparation and post-processing strategies, have led to additional gains in accuracy and robustness.\cite{Shirakawa2025}

Moreover, the effectiveness of QSCI critically depends on the quality of the prepared quantum state: the ansatz must exhibit a sufficiently large overlap with the exact eigenstate to ensure that the sampled configurations meaningfully represent the dominant contributions to the true wavefunction.
If this overlap is small, the sampling procedure becomes inefficient and the advantage over classical approaches is substantially diminished.
Physically motivated ans\"atze, such as unitary CC (UCC),\cite{Anand2022} require very deep circuits, making them impractical for larger systems. As an alternative, the unitary cluster Jastrow (UCJ) ansatz\cite{Matsuzawa2020} has been employed in QSCI implementations, in particular a local version that adapts to the connectivity of the quantum chip, known as the local unitary cluster Jastrow (LUCJ) ansatz.\cite{Motta2023} 
The parameters for this ansatz are typically obtained from a classical CC singles and doubles (CCSD) calculation, although recent work has explored improved parameter initialization and optimization strategies to better approximate the original CCSD amplitudes.\cite{Lin2025}

Another critical aspect in the practical application of QSCI is configuration recovery, a classical subroutine that corrects samples from the quantum computer that are corrupted by noise. Many of the samples exhibit incorrect Hamming weights, which can significantly disrupt the calculation if left uncorrected.

In this work, we assess the performance of the LUCJ ansatz as the state-preparation strategy in QSCI using ideal (noiseless) simulations to establish a reference baseline.
We then analyze the impact of quantum noise on the sampling process and its consequences for the resulting CI expansions.
Our results show that the combination of noise and configuration recovery can partially overcome the intrinsic limitations of the sampled LUCJ state by promoting a broader exploration of Hilbert space.
This behavior is examined by comparing noiseless sampling, simulations with noise models of varying strength, and data obtained from quantum hardware.
The critical role of configuration recovery is further illustrated through comparisons with random determinant selection.
All findings are demonstrated along the \ce{N2} molecular dissociation curve. 

\section{Methods and computational details} \label{sec:methods}
One and two electron integrals were obtained from the GitHub repository associated with Refs.~\citenum{RobledoMoreno2025} and  \citenum{jrobledo_github}. 
Unless explicitly stated, the circuits used in this work are truncated two-layer (2L') LUCJ circuits, identical to those employed in Ref.~\citenum{RobledoMoreno2025}.\cite{note-truncated-LUCJ} 
This choice enables a direct comparison between the quantum results reported in Ref.~\citenum{RobledoMoreno2025} and the exact simulations presented here, thereby supporting the conclusions drawn in this study.
To initialize the LUCJ parameters, a CCSD calculation was performed using the PySCF software package.\cite{pyscf}
The LUCJ quantum circuits were generated using the ffsim library\cite{ffsim} and implemented using Qiskit.\cite{Qiskit}
Exact simulations of these circuits were carried out using an extension of the ORQA formalism.\cite{Broers2025,Broers2026} 
The samples were obtained by taking $10^8$ shots from the circuits for each interatomic distance.
The amount of configurations obtained for each interatomic distance, $|\chi|$, is described in Section S1 of the Supporting Information.

The sets of noisy samples, $|\chi_{noisy}|$, were generated from the exact set using Algorithm \ref{noise_alg} (details in Section~\ref{subsec:noise_introduction}). For the random sets, bitstrings were selected from a uniform distribution across the entire Fock space until the number of configurations matched that of the exact sets, ensuring $|\chi| = |\chi_{\text{rand.}}|$ for each distance.\\
Throughout this study, the batch strategy introduced in Ref.~\citenum{RobledoMoreno2025} was employed. Multiple batches of configurations $\mathcal{S}^{(k)}$ of dimension $d$ are constructed from the sets $|\chi|, |\chi_{noisy}|$ and $|\chi_{rand.}|$, where $k$ indicates the batch number, with $k=10$ used throughout. To construct each batch, configurations are sampled from these sets, identifying and adding the unique $\alpha$- and $\beta$-strings to the batch. Then, determinants formed from all possible combinations of these strings are built, ensuring spin conservation.

Two approaches were used to construct the batches. The first fixes the subspace size of the batches. In this approach, configurations are drawn from the sample sets until $\sqrt{d}$ distinct strings are obtained, yielding batches with a fixed subspace size of $d$ after recombination. For the $N_2$ dissociation studied in this work, $d = 16 \times 10^6$. This method is used in Sections~\ref{subsec:exact_sampling}, \ref{subsec:noisy_sampling}, and \ref{subsec:random_samplign}.\\
The second approach follows the procedure from Ref.~\citenum{RobledoMoreno2025}, where a fixed number of configurations $\sqrt{d}/2$ are drawn from the sample set, the unique $\alpha$-strings are identified (yielding at most $\sqrt{d}$ distinct strings), and a subspace of size $\leq d$ is created. Results obtained with this method are analyzed in Section~\ref{sec:fixed_shots}, enabling a more direct comparison with the results of Ref.~\citenum{RobledoMoreno2025}.\\
Once the batches are constructed, CI Hamiltonian diagonalization, denoted $\hat{H}_{\mathcal{S}^{(k)}}$, is performed on the Fugaku supercomputer using the code from Ref.~\citenum{sbd_github}. By considering multiple batches with distinct configurations, the algorithm increases the probability of capturing the most relevant determinants, leading to an improved approximation of the ground-state energy. In all simulations, the reported energy corresponds to the minimum energy obtained across $K=10$ batches, i.e., $\min_k(E^{(k)})$, following the same protocol as in the original work \cite{RobledoMoreno2025}.
The configuration recovery implementation used in this work follows Ref.~\citenum{RobledoMoreno2025} and is provided through the Qiskit software package \cite{Qiskit}. In all cases, five iterations were used for each application of the recovery procedure.


\section{Limitations of LUCJ ansatz in QSCI}

\subsection{The local unitary cluster Jastrow} 
The limitations of near-term quantum devices impose stringent constraints on the design of wavefunction ans\"atze.
They must be sufficiently expressive to capture the essential features of molecular electronic structure, while maintaining a shallow circuit depth to remain compatible with limited coherence times and gate fidelities.
These considerations have motivated the use of the LUCJ ansatz,\cite{Motta2023} a local variant of the UCJ ansatz.\cite{Matsuzawa2020}
The UCJ ansatz is defined as:
\begin{equation}
    |\Psi \rangle = \prod_{\mu=1}^L e^{\hat{K}_\mu}e^{i\hat{J}_\mu}e^{-\hat{K}_\mu} |\phi_0 \rangle
    \label{eq:ucj}
\end{equation}
where $|\phi_0\rangle$ is a reference configuration (typically the HF determinant) and $L$ denotes the number of product layers. 
The operators $\hat{K}_\mu$ and $\hat{J}_\mu$ are given by:
\begin{eqnarray}
    \hat K_\mu = \sum_{pq,\sigma} {\mathcal{K}^\mu_{p \sigma} \hat{a}_{p \sigma}^{\dagger}\hat{a}_{q \sigma}} \\ 
    \hat J_\mu = \sum_{pq,\sigma \tau} \mathcal{J}^\mu_{pq,\sigma\tau} 
    \hat n_{p\sigma} \hat n_{q\tau}
\end{eqnarray}
where $p,q$ label molecular spatial orbitals, and $\sigma,\tau \in \alpha,\beta$ denote spin projections.
The coefficients $\mathcal{K}^\mu_{pq,\sigma}$ form an anti-Hermitian matrix, generating orbital rotations, while $\mathcal{J}^\mu_{pq,\sigma\tau}$ is a real symmetric matrix encoding density-density correlations.
Here, $\hat{n}_{p \sigma} = \hat{a}_{p \sigma}^{\dagger}\hat{a}_{p \sigma}$ is the number operator.
This structure combines unitary orbital rotations with diagonal correlators in the occupation-number basis, yielding a flexible yet hardware-efficient ansatz.
In its local (LUCJ) formulation, additional sparsity constraints are imposed on the $\mathcal{J}$ tensor to reduce circuit depth while retaining the dominant short-range correlation effects, making it particularly suitable for implementation on near-term quantum processors.
Concretely, the local approximation restricts the range of the density-density correlators in $\hat{J}_\mu$, leading to simplified expressions for the same-spin (SS) and opposite-spin (OS) contributions:
\begin{eqnarray}
    \hat J_\mu^{(SS)} = \sum_p \mathcal{J}_{p(p+1),\sigma\sigma}^\mu \hat{n}_{p\sigma} \hat{n}_{(p+1)\sigma}\\
    \hat J_\mu^{(OS)} = \sum_p \mathcal{J}_{pp,\alpha\beta}^\mu \hat{n}_{p\alpha} \hat{n}_{p\beta}
\end{eqnarray}
In this form, same-spin correlations are limited to nearest-neighbor orbital pairs, while opposite-spin correlations are restricted to on-site interactions.
This locality assumption drastically reduces the number of variational parameters and the corresponding circuit depth, while still capturing the dominant short-range correlation effects that are most relevant for molecular systems in a localized orbital basis.
The implementation of the  orbital rotations $e^{\hat{K}_{\mu}}$ remains unaffected in the local version of the ansatz. 
In the following, we consider LUCJ ans\"atze with $L=1$ and $L=2$, and on the truncated two-layer form only including the $\hat{K}_2$ term in the second layer,
\begin{equation} \label{eq:l2_truncated}
|\Psi \rangle = e^{\hat{K}_2} e^{\hat{K}_1} e^{i\hat{J}_1} e^{-\hat{K}_1} |\phi_0 \rangle
\end{equation}
which is expected to provide a balanced compromise between wavefunction expressivity and manageable circuit depth.
Moreover, this choice facilitates direct comparison with previous studies,\cite{RobledoMoreno2025} where the same state-preparation strategy was employed.

\subsection{Exact sampling of LUCJ expansions}\label{subsec:exact_sampling}
To assess the capabilities of the different LUCJ ans\"atze, we benchmark the performance of QSCI combined with noiseless sampling of the three variants considered ($L=1$, $L=2$, and the truncated two-layer form in equation~\ref{eq:l2_truncated}) for the computation of the \ce{N2} potential energy curve.
The LUCJ parameters were obtained from CCSD $t_2$ amplitudes.
All simulations are carried out using the cc-pVDZ basis set, correlating 10 valence electrons while freezing the $1s$ core orbitals, which results in 26 active molecular orbitals.
The configurational space is generated by sampling occupation bitstrings from the prepared quantum state until 4000 distinct $\alpha$ strings are collected.
The corresponding set of $\beta$ strings is then constructed by duplicating the sampled $\alpha$ occupations, and the full set of Slater determinants is obtained by combining all possible $\alpha$–$\beta$ pairs.
This procedure yields up to $1.6 \times 10^7$ configurations in the CI expansion.
The classical diagonalization of the CI Hamiltonian is performed using a batch strategy, and the final QSCI energy is taken as the lowest eigenvalue obtained across all batches (details in Section~\ref{sec:methods}).
This strategy provides a practical compromise between the size of the selected configurational space and the feasibility of the classical post-processing step.

\begin{figure}[H]
    \centering
    \includegraphics[width=8.0cm]{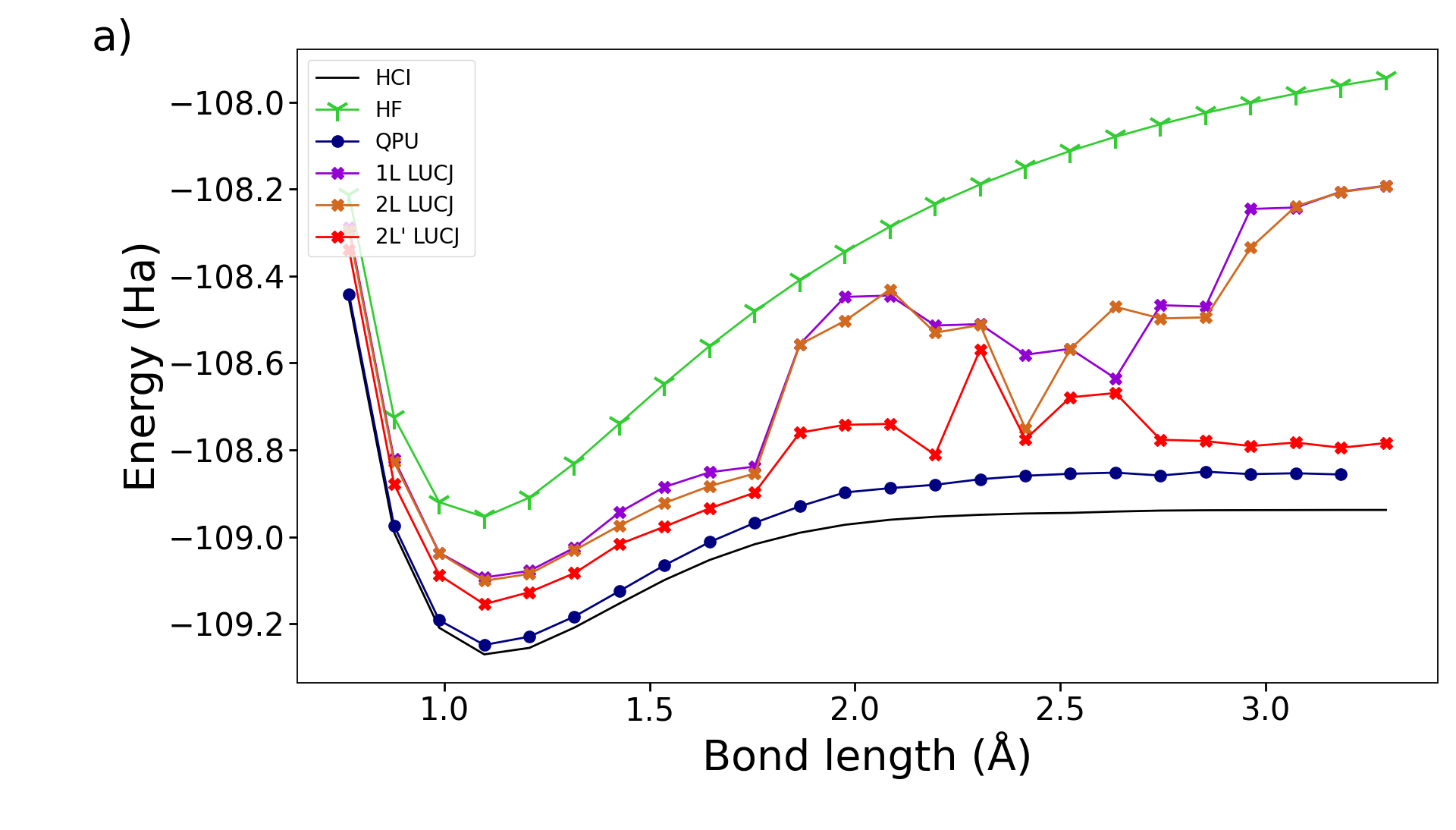}
    \includegraphics[width=5.0cm]{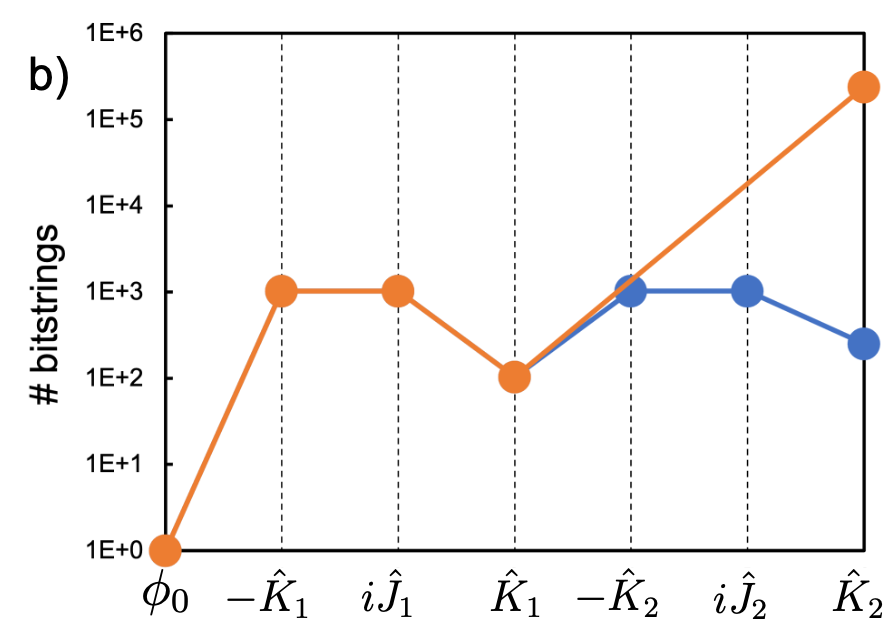}
    \caption{(a) QSCI energy profiles for the \ce{N2} dissociation obtained by sampling with cc-pVDZ basis and (10e, 26o). Violet, brown and red lines represent the exact circuit simulation for 1, 2 and truncated-2 (2L') layers respectively. Navy blue line are the data from the IBM experiment carried out in Ref.~\citenum{RobledoMoreno2025}.
    (b) Size of the complete CI space generated after the application of the different terms of LUCJ up to $L=2$ in equation~\ref{eq:ucj} (blue) and skipping $-\hat{K}_2$ and $i\hat{J}_2$, i.e., 2L', in equation~{\ref{eq:l2_truncated}} (orange),  obtained at 2.0~{\AA} of interatomic distance.}
    \label{fig:exact_n2_diss}
\end{figure}

In Figure~\ref{fig:exact_n2_diss}a, we present the energy profiles along the \ce{N2} dissociation obtained with QSCI using configurations sampled noiselessly from the different LUCJ ans\"atze, and compare them with the HF reference and the highly accurate HCI results. 
The QSCI energy curves obtained from the 1L and 2L LUCJ expansions are very similar to each other, yet both remain significantly higher in energy than the HCI reference.
These limitations become more pronounced at large interatomic separations, where the diagonalization of the generated CI spaces substantially overestimates the ground-state energy. 
In addition, the resulting potential energy curves are far from smooth, particularly in the stretched-bond region.
Interestingly, the results improve markedly when configurations are sampled from the truncated 2L LUCJ ansatz.
In this case, the QSCI energies are systematically lower across the entire range of bond distances and yield a significantly smoother potential energy profile.

To rationalize the improved performance of the truncated 2L ansatz (2L') compared with the 1L and full 2L forms, we analyze the size of the Hilbert space effectively spanned by each LUCJ expansion as a proxy for their expressivity (Figure~\ref{fig:exact_n2_diss}b).
Notably, while the 1L and 2L ans\"atze explore CI spaces on the order of $\sim10^2$ configurations, the 2L' form spans a space that is roughly three orders of magnitude larger.
This leads to the subspaces spanned by the 1L and 2L ans\"atze being significantly smaller than those obtained from the 2L' sampling.
While the 2L' LUCJ ansatz enables Hamiltonian diagonalization in subspaces containing up to $1.6\times10^{7}$ configurations (as previously indicated), the sampling of the 1L and 2L ans\"atze does not yield a sufficient number of configurations to construct subspaces of comparable size.
In these two cases the maximum number of available determinants has been used, generating subspaces of different sizes (see Support Information S2 for subspace size details).
As illustrated in Figure~\ref{fig:exact_n2_diss}b, the inclusion of the $\hat{K}_\mu$ operators that complete each $\mu$-th layer strongly concentrates the wavefunction weight around a small set of dominant configurations.
This compression significantly limits the diversity of determinants accessible through sampling, thereby reducing the effective configurational space explored by the ansatz.

Despite the improvements obtained when sampling from the truncated 2L ansatz, the resulting QSCI energies remain significantly higher than the HCI reference values, highlighting the intrinsic limitations of LUCJ as a sampling state within QSCI, as also suggested in previous studies.\cite{Reinholdt2025,Lin2025}
More strikingly, the energies obtained from ideal circuit simulations are noticeably worse than those reported from quantum hardware using the same truncated 2L ansatz, particularly at interatomic distances between 1.7 and 2.6~\AA.
Notice that, the subspace generated from noiseless sampling is larger that the one obtained from the QPU sampling in Ref.~\citenum{RobledoMoreno2025} (details in Section~\ref{sec:methods}). Larger subspaces have been used in Section S3 of the Supporting Information, where the \ce{N_2} dissociation was calculated using the full set of configurations, reaching subspaces up to $1.2\times10^{8}$. This is equivalent to performing the diagonalization of the full space spanned by the LUCJ circuit. The results do not improve the energy substantially, remaining very similar to those in Fig.~ \ref{fig:exact_n2_diss}.
This observation suggests that the presence of hardware noise, together with post-processing procedures for configuration recovery, may play a significant role in the improved results obtained on real devices. 
In the following, we investigate this hypothesis in detail.

\section{The role of noise and recovery in QSCI}

\subsection{Noisy sampling and recovery}

Sampling a prepared quantum state $\Psi$ using an $N$-qubit circuit yields a set $\tilde{\chi}$ of measured bitstrings $\mathbf{x} \in \{0\vee1\}^N$, drawn according to a probability distribution $\tilde{P}_{\Psi(\mathbf{x})}$ that reflects the weight of each configuration in the state $\Psi$,
\begin{equation}
\tilde{\chi} = \{ \mathbf{x} \mid \mathbf{x} \sim \tilde{P}_{\Psi(\mathbf{x})} \}.
\end{equation}
Here, the tilde notation in $\tilde{\chi}$ and $\tilde{P}_{\Psi}$ denotes that the sampling process is affected by noise.
In practical quantum devices, gate imperfections and readout errors distort the ideal probability distribution, leading to erroneous bitstrings and deviations from the noiseless sampling statistics.

To mitigate the effect of noise in the sampled set $\tilde{\chi}$, configuration recovery has been introduced as a classical post-processing procedure.\cite{RobledoMoreno2025}
This method consists of a self-consistent algorithm that is applied iteratively to the noisy samples in order to correct unphysical configurations.
In particular, configuration recovery targets the violation of particle-number conservation induced by noise. In the ideal sampling process, each configuration should preserve the total number of particles in the system.
For spin-$1/2$ particles such as electrons, this constraint further requires the conservation of the number of particles within each spin sector, namely the spin-$\alpha$ and spin-$\beta$ orbital subspaces.
However, hardware noise may flip qubits during the circuit execution or measurement, generating bitstrings that do not satisfy these physical constraints.
Configuration recovery identifies and corrects such bitstrings to restore the appropriate particle-number and spin-sector occupations.

In the first iteration of the configuration recovery procedure, the samples in $\tilde{\chi}$ that already satisfy the correct number of particles in each spin sector are used to construct the initial CI subspace.
The electron occupation of each spin-orbital is then computed from the lowest eigenstate of the corresponding Hamiltonian or, when the CI problem is solved in batches (see details in Section~\ref{sec:methods}), as the average over the batch ground states,
\begin{equation}\label{occ_calc}
n_{p\sigma} = \frac{1}{N_b}\sum_{k=1}^{N_b} \langle \Psi^{(k)} | \hat{n}_{p\sigma} | \Psi^{(k)} \rangle,
\end{equation}
where $\{\Psi^{(k)}\}$ are the lowest-energy eigenstates obtained for the $N_b$ batches.
Next, the occupation strings $\mathbf{x}$ (bitstrings) that contain an incorrect number of $\alpha$ and/or $\beta$ electrons are corrected by performing the required bit flips.
The bits to be flipped in each spin sector are selected probabilistically according to a function (see Ref.~\citenum{RobledoMoreno2025}) that depends on the deviation between the current bit value and the corresponding average orbital occupation, $|x_{p\sigma} - n_{p\sigma}|$, where $x_{p\sigma}$ denotes the occupation (0 or 1) of the spin-orbital $p\sigma$ in the bitstring $\mathbf{x}$.
This procedure generates a new set of corrected configurations that satisfy the particle-number constraints, which can then be used to construct an improved CI space.
Diagonalization within this updated space yields refined ground states and updated orbital occupations $\{n_{p\sigma}\}$, which serve as input for the next recovery iteration.
Importantly, each recovery step is applied to the original noisy sample set $\tilde{\chi}$.

\subsection{Sampling noise model}\label{subsec:noise_introduction}
To investigate the impact of sampling noise and configuration recovery, we introduce a noise model that allows us to incorporate controlled errors into the samples of LUCJ ans\"atze obtained from our simulations based on an extension of the ORQA formalism.\cite{Broers2025,Broers2026}
This approach enables a systematic analysis of how different noise levels affect the sampled configuration space and the resulting QSCI energies.

The model is intentionally simple and is defined by a single parameter, denoted by $p$.
For each ``parent'' bitstring in the original (noiseless) sample set, a randomly selected bit is flipped with probability $p$. 
If a flip occurs, an additional random bit may be flipped with the same probability, and this process is repeated until no further flips occur. 
The resulting ``descendant'' bitstring is then added to the sample set with the same number of counts as the ``parent'' configuration.
In the limit $p=1$, each parent bitstring generates a descendant with identical counts, effectively doubling the total number of samples, although different descendants may coincide by chance.
For intermediate values of $p$, the number of distinct configurations can increase by at most a factor of two. By construction, the total number of counts is always doubled, while the original configurations remain present in the noisy sample set.
An outline of the noise-model workflow is provided in Algorithm~\ref{noise_alg}.
\begin{algorithm}[H]
\caption{Noise model algorithm}\label{noise_alg}
\begin{algorithmic}[1]
\For{$parent$ bitstring $b$ in samples}
    \State $descendant \gets b$ \Comment{Start with a copy of the parent}
    \While{True}
        \State generate random number $r \in [0,1]$
        \If{$r < p$}
            \State select random bit in $descendant$ and flip it
        \Else
            \State \textbf{break} \Comment{If no flip, move to next parent}
        \EndIf
    \EndWhile
    \State add $descendant$ with same counts as $b$
    \If{no flips occurred}
        \State double the counts of $b$
    \EndIf
\EndFor
\end{algorithmic}
\end{algorithm}

\subsection{Noisy sampling of LUCJ}\label{subsec:noisy_sampling}
\begin{figure}[H]
    \centering
    \begin{subfigure}{0.48\textwidth}
        \centering
        \includegraphics[width=\linewidth]{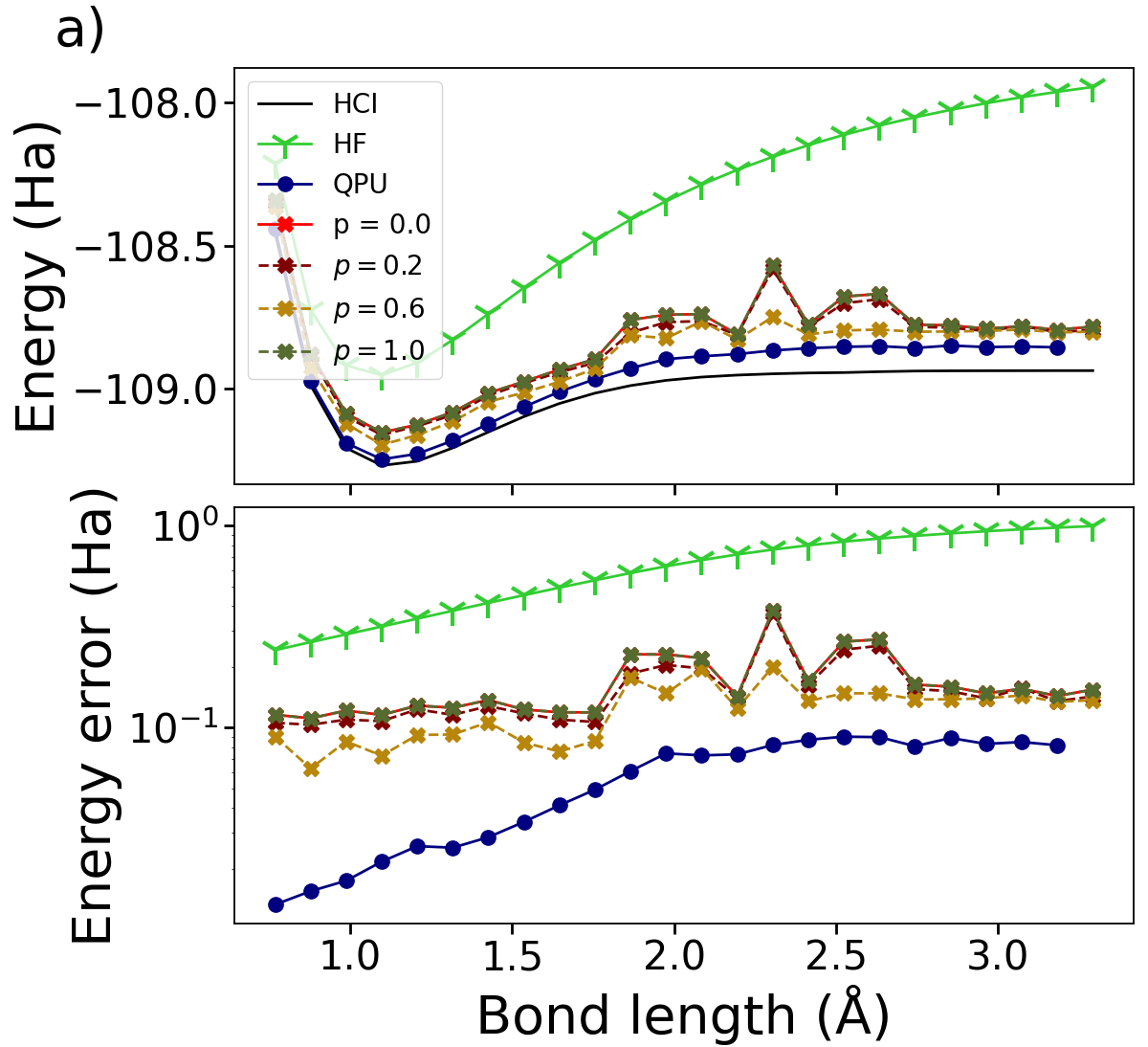}
        \caption{Recovery iteration 0}
        \label{fig:noisy_recit0}
    \end{subfigure}
    \hfill
    \begin{subfigure}{0.48\textwidth}
        \centering
        \includegraphics[width=\linewidth]{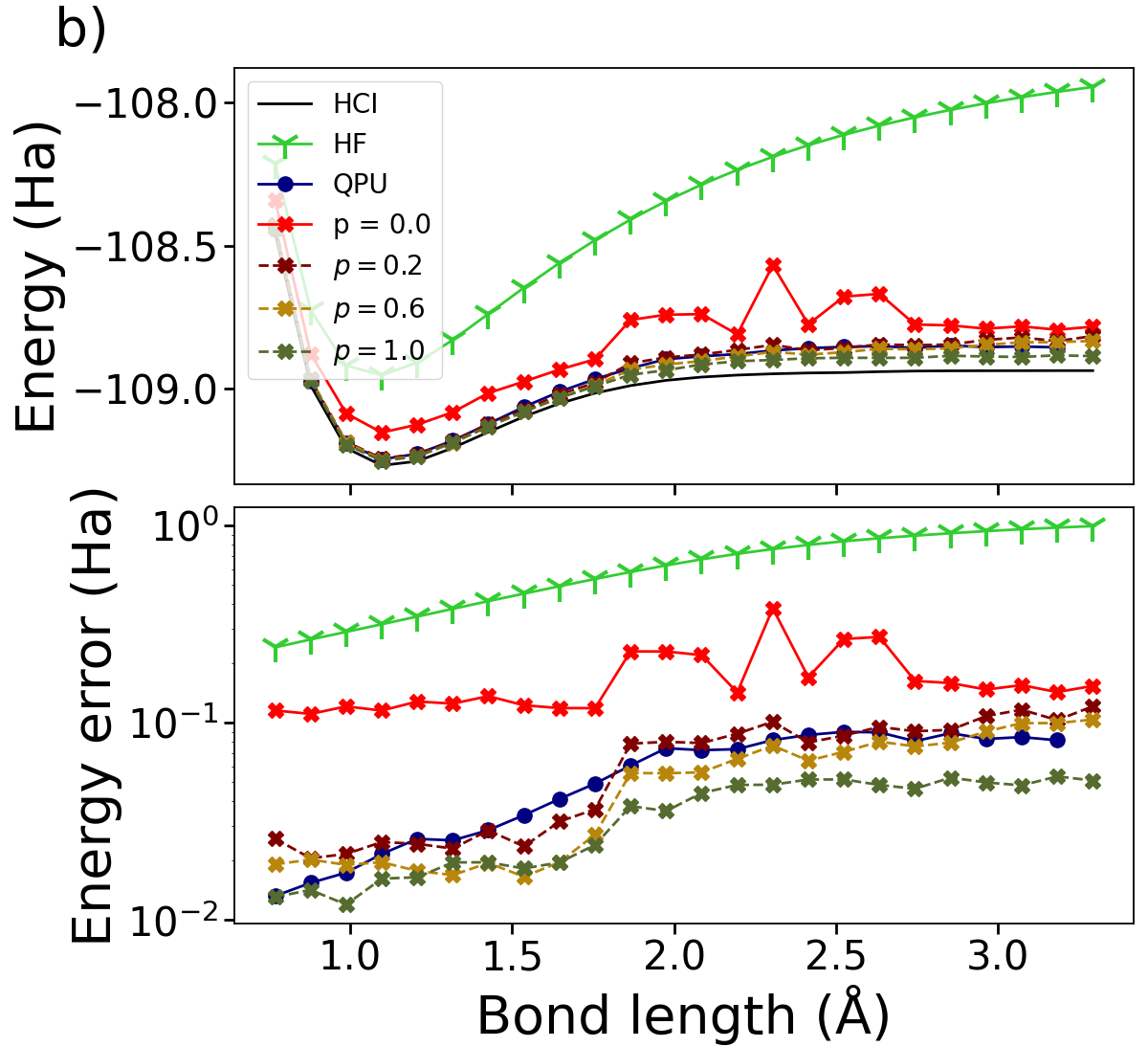}
        \caption{Recovery iteration 5}
        \label{fig:noisy_recit4}
    \end{subfigure}

    \vspace{1em}

    \caption{$N_{2}$ dissociation with cc-pVDZ basis and (10e, 26o).
    Maroon, gold and green lines represent exact results with different noise levels, from $p=0.2$ (yellow) to $p=1$ (red). The red line represents ORQA noiseless exact results. Figures (a) and (b) show the energies (up) and energy errors with respect to HCI (down) during the dissociation  for configuration recovery iterations 0 and 5 respectively.}
    \label{fig:noisy_recits_0_4}
\end{figure}

The noise model described above was applied to the calculation of the ground-state energy profile along the \ce{N2} dissociation for different noise levels. For each noisy sample set, five iterations of the configuration recovery procedure were subsequently performed to correct the sampled configurations. Figure~\ref{fig:noisy_recits_0_4} compares the results obtained by introducing noise levels of $p = 0.2$, $0.6$, and $1.0$ into the exact sampling distribution with those from ideal (noiseless) sampling, sampling performed on real quantum hardware,\cite{RobledoMoreno2025} and the HCI reference energies.

Before applying configuration recovery, the CI space generated by the noisy sampling model consists of the configurations obtained from exact (noiseless) sampling together with additional noisy bitstrings. A fraction of these bitstrings must be discarded because they do not preserve the correct Hamming weight, i.e., they do not contain the proper number of electrons in each spin sector.
For the highest noise level ($p = 1$), nearly all noisy samples fall outside the target subspace. 
As a consequence, most of them are discarded due to an incorrect Hamming weight, leaving essentially only the original exact LUCJ samples.
This results in an energy curve that closely matches that obtained from noiseless LUCJ sampling (Figure~\ref{fig:noisy_recit0}).
In contrast, lower noise levels ($p = 0.2$ and $p = 0.6$) lead to slightly lower energies than in the exact case, as the noise model can generate additional bitstrings with the correct Hamming weight in each spin sector that were not present in the original exact sampling distribution.
Despite this moderate improvement, the results still fall short of the performance achieved with sampling on real quantum hardware combined with configuration recovery.

On the other hand, the application of the iterative configuration recovery procedure markedly improves the dissociation energy profile, which becomes smooth and yields energies of comparable quality (at least) to those obtained from quantum hardware sampling (Figure~\ref{fig:noisy_recit4}).
Notably, increasing the noise level leads to a reduction in the energy errors, highlighting the importance of the combined effects of noise and configuration recovery in overcoming the limitations of the LUCJ prepared state within QSCI.

\subsection{Random selection}\label{subsec:random_samplign}
To further illustrate the critical role of the configuration recovery procedure, we consider the limiting case in which no physically motivated configuration selection is employed.
Instead, configurations are generated through a completely random process, drawn from a uniform distribution over the full Fock space of $N$-bit strings.
From this pool, we select the same number of configurations as obtained from the exact sampling of the LUCJ state.
To incorporate a minimal degree of chemical intuition, we additionally include the mean-field reference configuration, namely the HF determinant, in the selected set.

\begin{figure}[H]
    \centering
    \includegraphics[width=12cm]{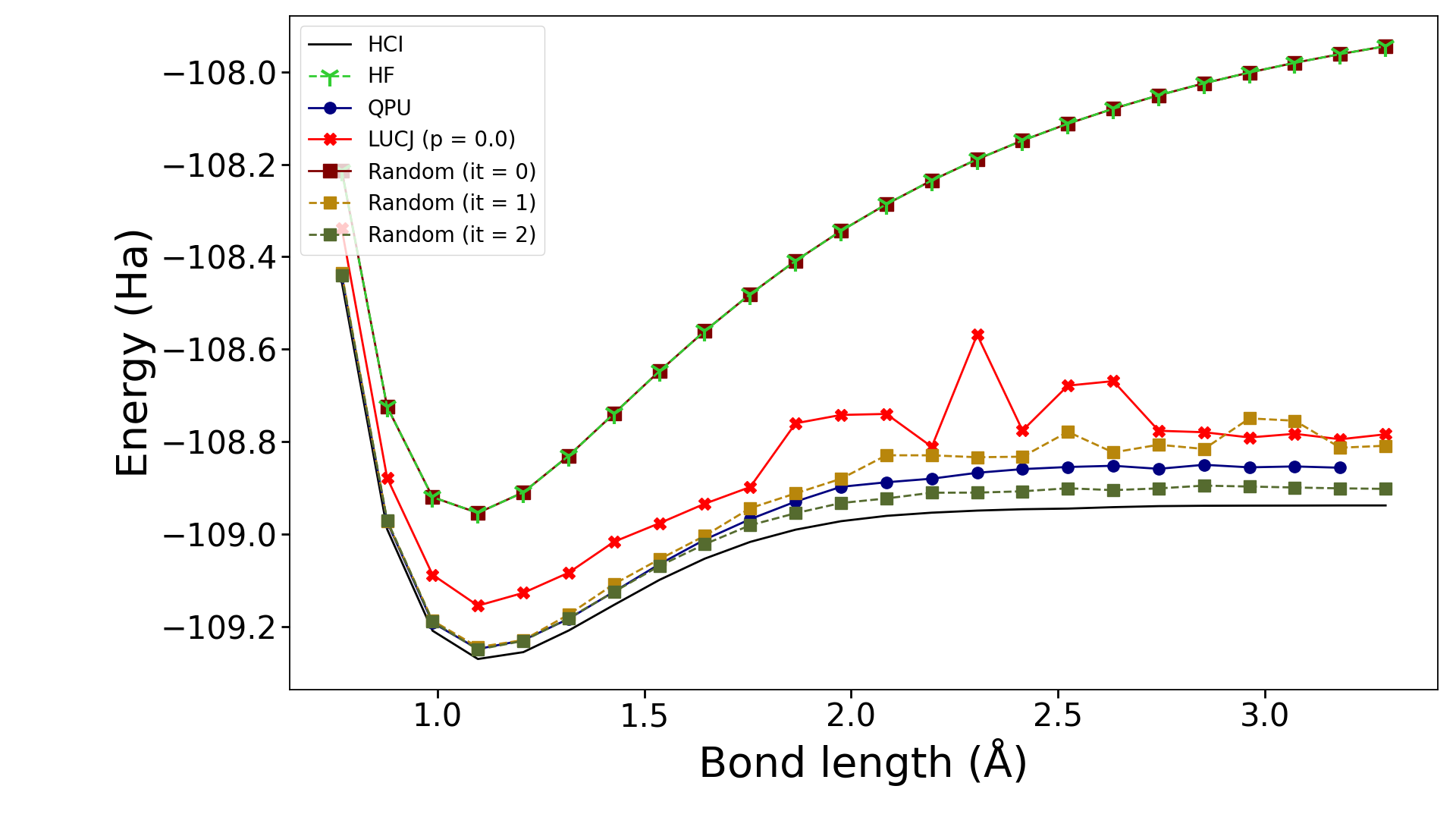}
    \caption{\ce{N2} dissociation with cc-pVDZ basis and (10e, 26o).
    Maroon, gold and green lines represent the energy obtained using samples drawn from a uniform distribution at different configuration recovery iterations, 0, 1 and 2 respectively. For each distance, the number of configurations generated from the uniform distribution matches the number of configurations obtained from the exact simulation, plus the HF configuration.}
    \label{fig:unif_recits_0_1_2}
\end{figure}

The CI energy profiles for the \ce{N2} dissociation obtained from randomly generated bitstrings are shown in Figure~\ref{fig:unif_recits_0_1_2}, together with the results after one and two configuration recovery iterations. 
These curves are compared with the HF energy, the energies obtained from noiseless and quantum-hardware sampling of the LUCJ ansatz, and the HCI reference.
The results show that the energies obtained from the raw randomly generated occupation strings (without recovery) essentially coincide with the mean-field results.
This behavior arises because the random selection of bitstrings from the full Fock space of 10 electrons in 26 orbitals rarely produces configurations with the correct Hamming weight, and therefore almost no additional determinants contribute to the CI expansion.
Remarkably, the application of just a single configuration recovery iteration leads to a substantial improvement of the CI energy profile.
The resulting curve is already smoother and yields lower energies than those obtained from exact LUCJ sampling. 
After the second iteration, once a sufficiently accurate set of orbital occupations has been established, the energies approach the HCI reference values and even surpass those obtained from sampling on real quantum hardware.

This result demonstrates that, with only minimal chemical input, i.e., the inclusion of the HF determinant, iterative configuration recovery can rapidly overcome the limitations of the LUCJ ansatz and achieve energies of at least the same quality as those obtained from sampling on quantum hardware. In fact, by recovery iteration 2, the energies surpass those obtained from the QPU. This difference is attributed to the  batch construction method used (see details in Section~\ref{sec:methods}). A comparison using an identical batch construction method is discussed in Section~\ref{sec:fixed_shots}.
These findings further suggest that the performance observed in hardware-based QSCI calculations with LUCJ-prepared states is largely attributable to the configuration recovery procedure. 

\section{Size of the configurational space}\label{sec:fixed_shots}
A detailed inspection of the energy curves in Figure~\ref{fig:unif_recits_0_1_2} reveals that randomly generated configurations combined with configuration recovery can yield lower energies than those obtained from sampling the LUCJ ansatz on real quantum hardware.
At first sight, this observation may appear surprising.
However, it is worth noting that, for the \ce{N2} system at the same level of theory (10 electrons in 26 orbitals), the fraction of sampled configurations with the correct electron number obtained from real quantum hardware amounts to only about 0.17\% of the total generated samples.\cite{RobledoMoreno2025}

While the overall performance can largely be attributed to the configuration recovery procedure, there is no intrinsic reason why highly noisy quantum sampling should lead to worse results than those obtained from an initially uniform distribution.
The apparent discrepancy instead arises from differences in the effective size of the CI space explored in each case.
In the real quantum experiment, the dimension of the CI space is ultimately limited by the number of selected determinants (i.e., 2000).
In contrast, in our simulations the LUCJ state (or the uniform distribution) is sampled until 4000 distinct $\alpha$ occupation strings are generated (see Section~\ref{sec:methods} for details).
Although the theoretical upper bound in both cases corresponds to 16 million possible configurations, our procedure typically produces a larger number of distinct determinants, which explains the lower energies obtained in the simulations.

\begin{figure}[H]
    \centering
    \includegraphics[width=8cm]{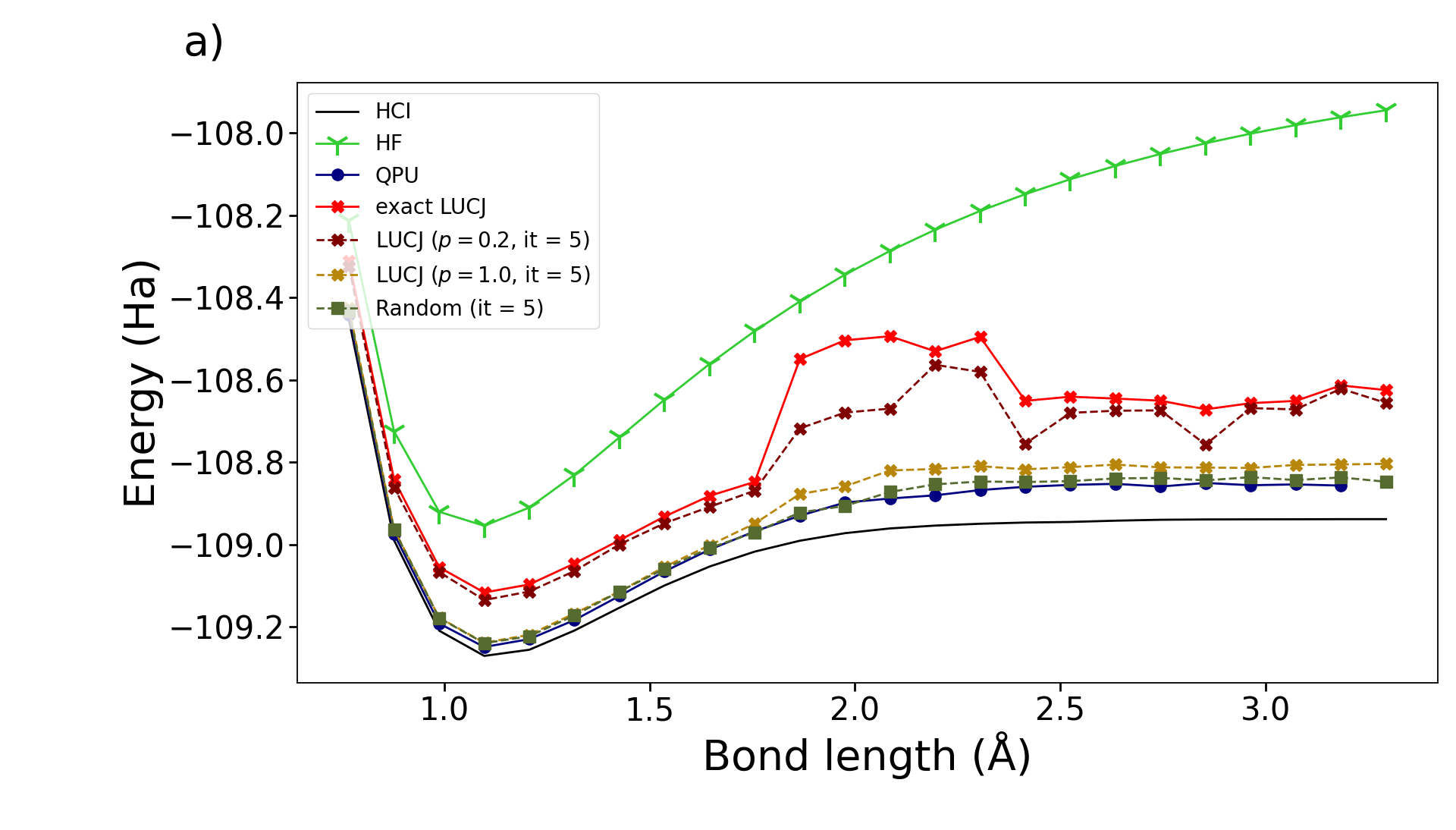}
    \includegraphics[width=8cm]{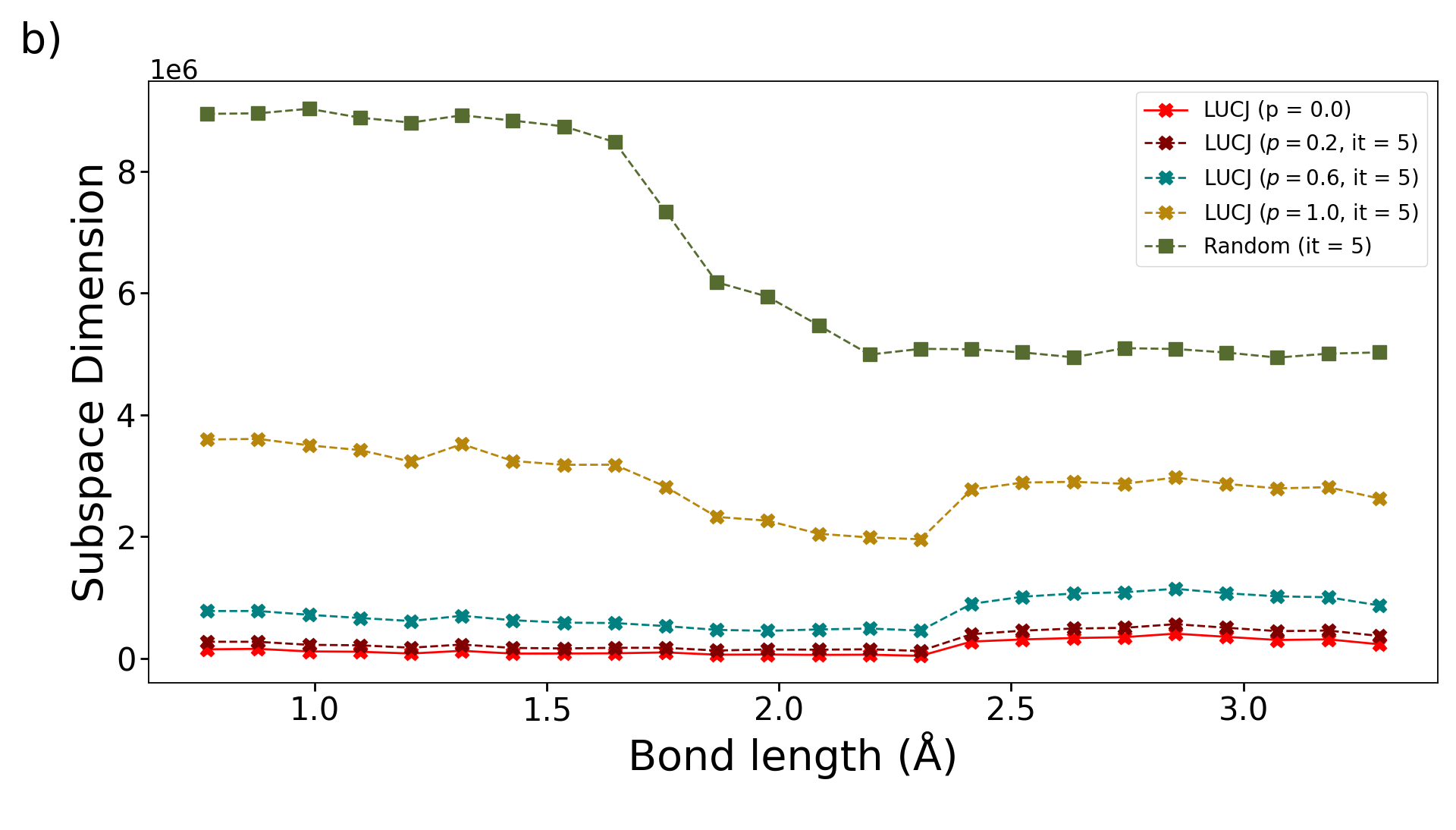}
    \caption{Energy (a) and size of the generated CI space (b) along the \ce{N2} dissociation with cc-pVDZ basis and (10e, 26o) with subspace size determined by fixed number of collected determinants (i.e., 2000). Energies from noisy samples and uniform distribution are obtained from the recovered set of samples applying 5 recovery iteration steps.}
    \label{fig:ener_cteshots}
\end{figure}

To enable a more direct comparison, we therefore repeat the simulations using exact and model-noise sampling of the truncated 2L LUCJ ansatz, as well as random configuration selection, while fixing the number of  collected determinants to 2000 (Figure~\ref{fig:ener_cteshots}). 
Beyond ensuring a fair comparison, this additional set of calculations also allows us to examine how the configuration recovery procedure applied to noisy samples influences the effective size of the explored Hilbert space.

The results obtained display behavior very similar to that discussed previously.
Exact LUCJ sampling significantly overestimates the energies and produces a rather poor dissociation profile, whereas higher noise levels combined with configuration recovery systematically improve the energy curves.
Notably, the results obtained from random configuration selection now nearly overlap with those from real quantum hardware sampling.

A clearer picture emerges when analyzing the number of configurations generated in each case (Figure~\ref{fig:ener_cteshots}b).
We observe a strong correlation between the number of recovered configurations and the resulting CI energies: higher noise levels enable the recovery procedure to generate a larger number of non-equivalent configurations beyond those produced by the pristine LUCJ circuit, leading to improved variational energies.
In other words, these results clearly demonstrate that the interplay between noise and configuration recovery effectively enhances the expressivity of the generated CI wavefunction, allowing significantly better solutions than those obtained from exact sampling of the LUCJ state alone.

\section{Conclusions} \label{sec:conclusions}

In this work, we have analyzed the performance of QSCI in combination with LUCJ ans\"atze, with particular emphasis on the interplay between ansatz expressivity, sampling, noise, and configuration recovery.
To this end, we have considered the dissociation of \ce{N2} using a relatively large active space, enabling access to an extended Hilbert space while simultaneously probing different correlation regimes, from equilibrium to the strongly correlated limit associated with triple-bond breaking.

Our results highlight intrinsic limitations of the LUCJ ansatz as a sampling state within QSCI.
Our analysis uncovers that the configurational spaces generated through noiseless sampling remain relatively small and highly concentrated around a limited set of dominant determinants. 
As a consequence, the resulting CI expansions lack sufficient diversity to accurately describe strongly correlated regimes, leading to significant overestimation of the ground-state energy, especially at stretched geometries.
These results highlight the need for improved ans\"atze in QSCI, either through an improved parameter initialization, or by completely replacing the LUCJ approach.
By introducing a controlled noise model, we have shown that noisy sampling can partially alleviate these limitations.
Moderate levels of noise broaden the sampling distribution and enable the generation of additional configurations beyond those supported by the ideal LUCJ state. When combined with configuration recovery, this leads to systematically improved energy profiles.
These findings indicate that noise, often regarded as detrimental, can in this context play a constructive role by enhancing the exploration of Hilbert space.
The critical importance of configuration recovery has been demonstrated through the limiting case of random configuration selection. Starting from a uniform distribution over the Fock space and including only minimal chemical input, iterative recovery alone is able to rapidly build physically meaningful CI spaces and achieve energies comparable to (or even better than) those obtained from quantum hardware sampling.
This clearly establishes configuration recovery as a key component of QSCI in current quantum hardware sampling, largely responsible for transforming otherwise unstructured or noisy samples into accurate wavefunction representations.
Finally, our analysis reveals that the overall performance of QSCI is strongly correlated with the effective size of the generated configurational space.
In this regard, both noise and recovery act as mechanisms that enhance the accessible Hilbert space, effectively increasing the expressivity of the resulting wavefunction beyond that of the original ansatz.

These results suggest that, within the QSCI framework, the combination of sampling, noise, and recovery may be more critical than the choice of the initial ansatz itself. This perspective opens new directions for the design of quantum algorithms, where controlled exploration of Hilbert space (rather than solely improving ansatz fidelity) may provide a more effective route toward accurate electronic structure calculations on near-term quantum devices.

\section*{Data and Software Availability}
The data supporting the findings of this study are available within the article and its Supporting Information, and have been made publicly available through the Zenodo repository.\cite{vaquero_sabater_2026_20269073} The code developed for circuit generation and Hamiltonian diagonalization is openly available on GitHub.\cite{VQEmulti}

\begin{acknowledgement}

We acknowledge financial support from MICIU/AEI/10.13039/501100011033 (project PID2022-136231NB-I00) and by FEDER, UE.
We also thank the support by the IKUR Strategy under the collaboration agreement between Ikerbasque Foundation and DIPC on behalf of the Department of Education of the Basque Government.
The authors are thankful for the technical and human support provided by the Donostia International Physics Center (DIPC) Computer Center.
Research at RIKEN was supported by the Japan Society for the Promotion of Science (JSPS) KAKENHI (Grant Nos. JP26K06972 and JP21H04446); the New Energy and Industrial Technology Development Organization (NEDO), Japan (Project No. JPNP20017); the Japan Science and Technology Agency (JST) through the COI-NEXT program (Grant No. JPMJPF2221); and the Ministry of Education, Culture, Sports, Science and Technology (MEXT), Japan, through the Program for Promoting Research on the Supercomputer Fugaku (Grant No. MXP1020230411). Additional support was provided by the UTokyo Quantum Initiative, the RIKEN TRIP initiative (RIKEN Quantum), and the Center of Excellence (COE) Research Grant in Computational Science from Hyogo Prefecture and Kobe City through the Foundation for Computational Science.

\end{acknowledgement}

\begin{suppinfo}
In the supportive information section further details about the theory and methodology are provided. These include...
\end{suppinfo}

\bibliography{abbr, references} 

\end{document}


\newpage
\tableofcontents
\clearpage
\section{Truncated Two-Layer LUCJ Subspace Dimension}
\begin{figure}[H]
    \includegraphics[width=0.7\textwidth]{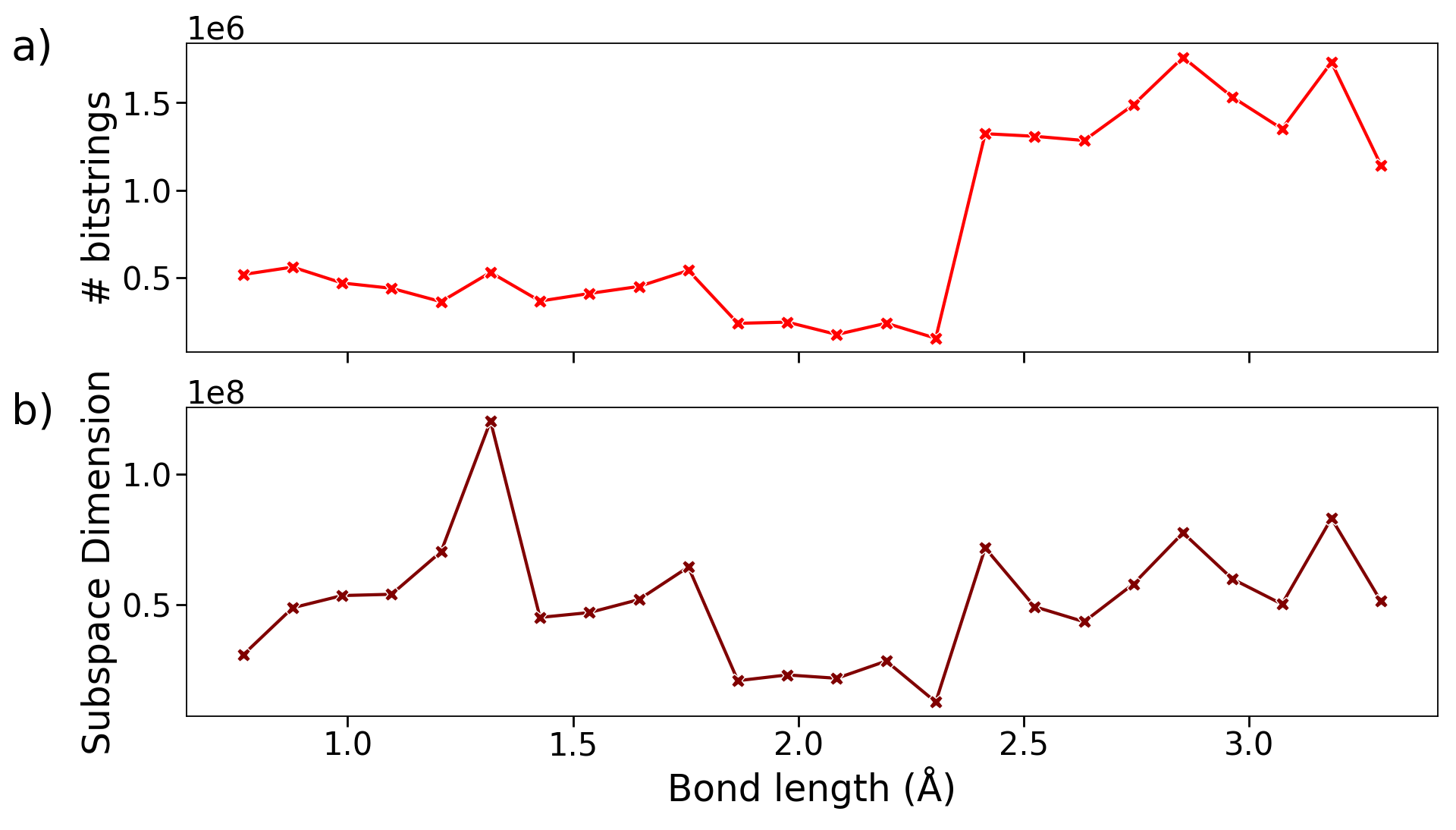}
    \caption{(a) Total number of determinants (bitstrings) obtained for each distance in \ce{N2} dissociation with cc-pVDZ basis and (10e, 26o). (b) Dimension of the subspace generated by the unique $\alpha$-occupations obtained from the sets of bitstrings in (a).}
    \label{fig_si:full_dim_trunclucj}
\end{figure}

As explained in Section 2 of the main text, for each interatomic distance, an exact LUCJ simulation with $10^8$ shots has been performed with ORQA, obtaining a probability distribution for the sampled bitstrings.
Assuming that almost all determinants present in the truncated two-layer LUCJ ansatz are measured after performing $10^8$ shots, Figure \ref{fig_si:full_dim_trunclucj} shows the dimension of the subspace spanned by this ansatz for each interatomic distance along the \ce{N2} dissociation.

Figure \ref{fig_si:full_dim_trunclucj}a shows the total number of different determinants sampled.
Figure \ref{fig_si:full_dim_trunclucj}b plots the subspace dimension generated as follows: first, identify all unique $\alpha$-occupations, second, construct the $\beta$-occupations by duplicating the $\alpha$ set, and third, create the complete set of Slater determinants by combining all possible $\alpha$-$\beta$ pairs.

\newpage

\section{One and Two-Layer LUCJ Subspace Dimension}

\begin{figure}[H]
    \centering

    \begin{subfigure}{0.48\textwidth}
        \centering
        \includegraphics[width=\linewidth]{figures/exact_lucj_diss_b.png}
        \label{fig_si:energies_exact}
    \end{subfigure}
    \hfill
    \begin{subfigure}{0.47\textwidth}
        \centering
        \includegraphics[width=\linewidth]{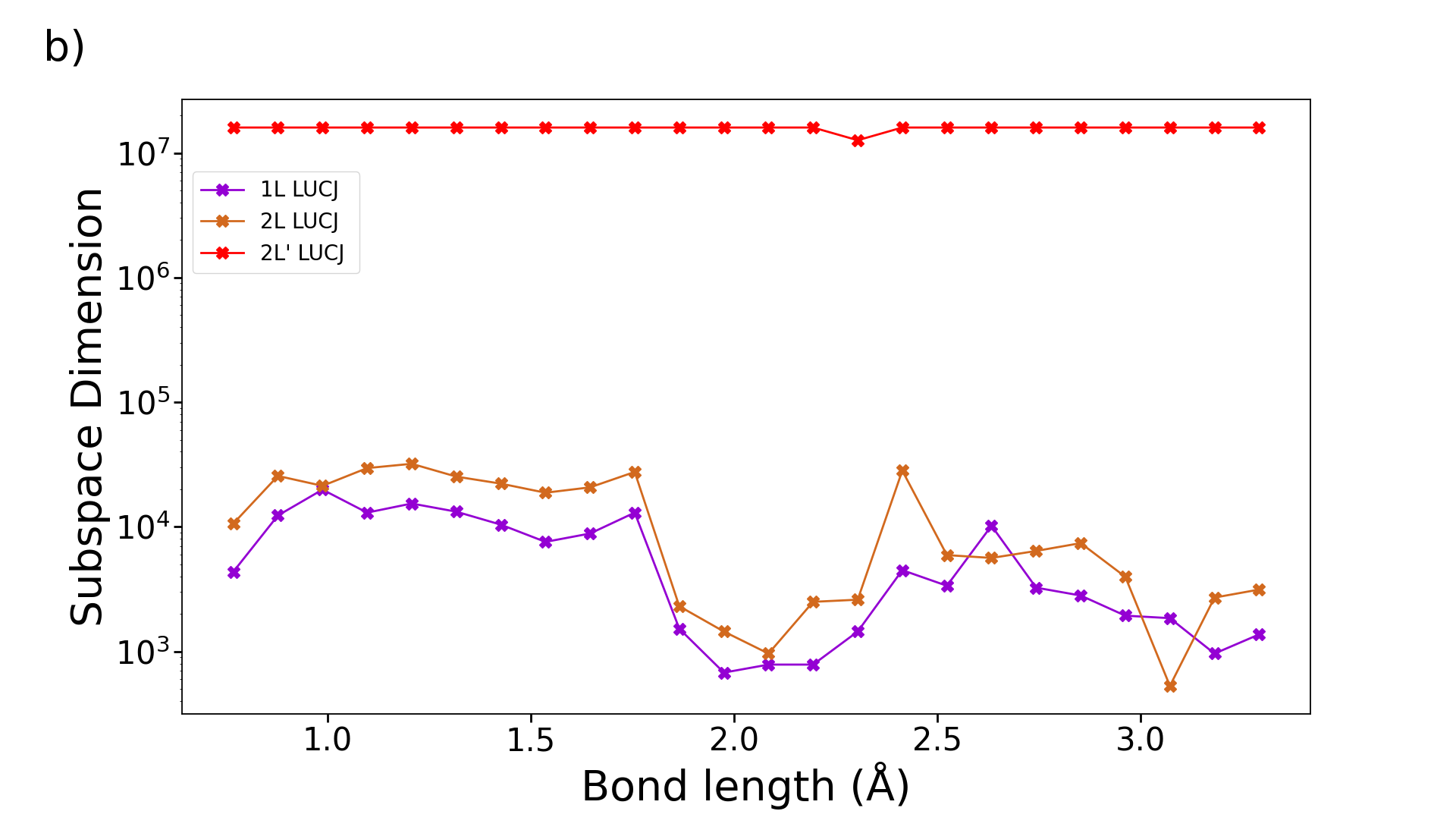}
        \label{fig_si:subs_exact}
    \end{subfigure}

    \caption{(a) QSCI energy profiles for the \ce{N2} dissociation obtained by sampling with cc-pVDZ basis and (10e, 26o). Violet, brown and red lines represent the exact circuit simulation carried out with ORQA for 1, 2 and truncated-2 layers respectively. Navy blue line are the data from the IBM experiment carried out in Ref.~\citenum{RobledoMoreno2025}.
    (b) Subspace dimension at each point of the \ce{N2} dissociation for 1, 2 and truncated-2 layers of LUCJ ansatz.}
    \label{fig_si:dims_sub_exact}
\end{figure}

As mentioned in the main text, subspaces for the Hamiltonian diagonalization are constructed by sampling bitstrings from the ansatz until 4000 unique $\alpha$ strings are collected. Then, the set of $\beta$ strings is constructed by duplicating the sampled $\alpha$ occupations, and the full set of Slater determinants is obtained by combining all possible $\alpha$–$\beta$ pairs. This yields subspaces of $1.6 \times 10^7$ configurations (red line in Figure~\ref{fig_si:dims_sub_exact}\subref{fig_si:subs_exact}b). 
However, as shown in Figure 1b, this is only possible when using the truncated two-layer ansatz, since sampling the non-truncated 1L and 2L does not generate enough unique $\alpha$ occupations to reach 4000.

When the coefficients in $e^{i\hat{J}_\mu}$ (coming from the CCSD amplitudes) are small, the addition of the rotation that ``closes'' the layer, i.e., $e^{-\hat{K}_\mu}$, approximately cancels the term $e^{\hat{K}_\mu}$. 
Therefore, the action of $e^{-\hat{K}_\mu}$ concentrates the wavefunction weight around a small set of dominant configurations.
In these two cases (1L and 2L LUCJ), where the amount of determinants does not reach 4000, all the configurations available have been used, generating subspaces with the maximum size possible. The dimension of these subspaces for both ansätze and each interatomic distance considered during the \ce{N2} dissociation is shown in Figure~\ref{fig_si:dims_sub_exact}\subref{fig_si:subs_exact}b.

\section{Full space diagonalization}
This section compares the energy profile for the same truncated two-layer ansatz using two different subspace sizes, the same 16 million-dimensional subspace used in the main text, and the maximum available subspace size, which corresponds to that from Figure~\ref{fig_si:full_dim_trunclucj}b.

\begin{figure}[H]
    \includegraphics[width=0.7\textwidth]{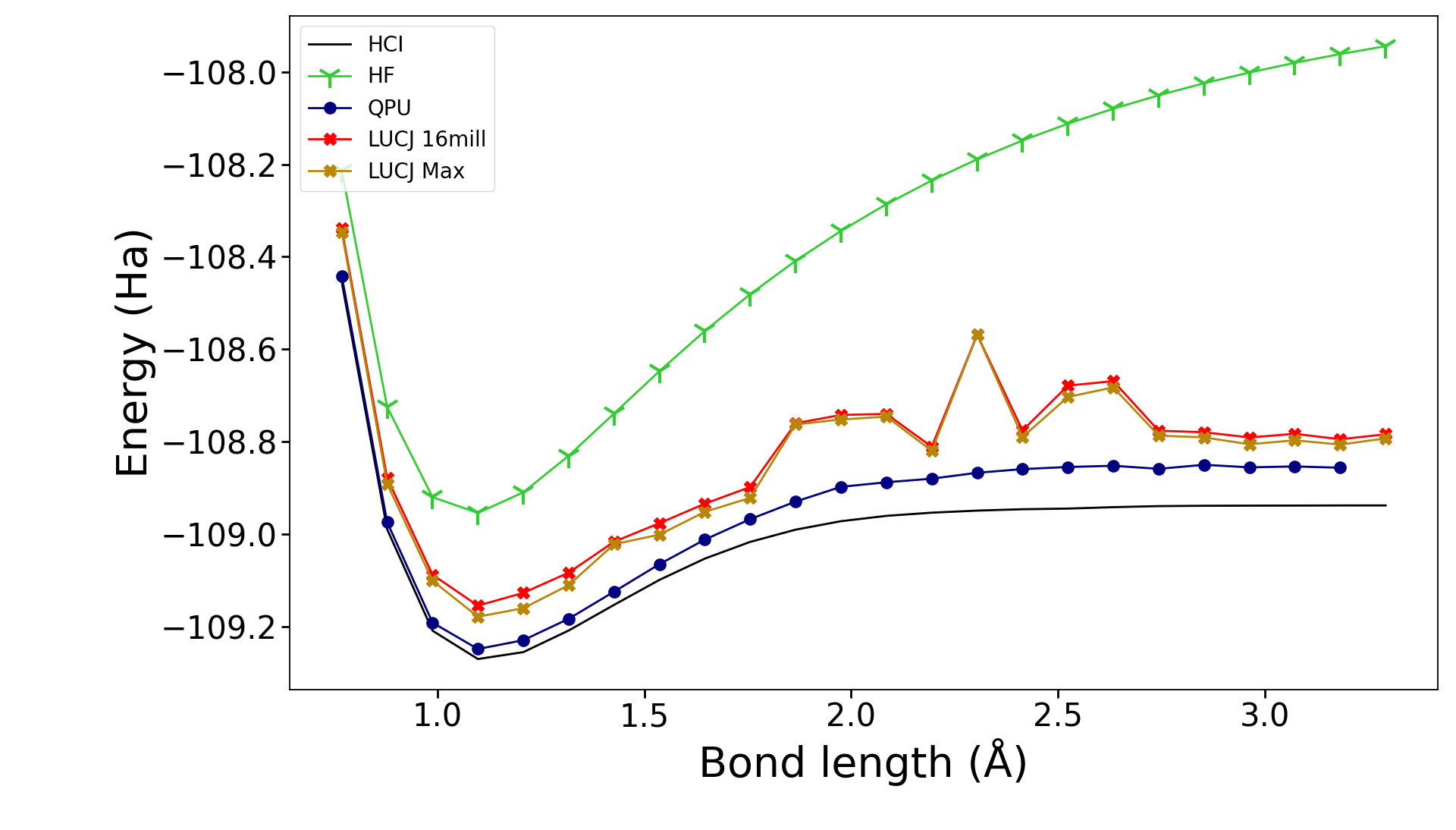}
    \caption{QSCI energy profiles for the \ce{N2} dissociation obtained by sampling with cc-pVDZ basis and (10e, 26o). Gold and red lines represent the exact circuit simulation carried out with ORQA for truncated two-layer ansatz using subspaces of $1.7\times 10^{7}$ and full subspace respectively. Navy blue line represents the data from the IBM experiment carried out in Ref.~\citenum{RobledoMoreno2025}.}
    \label{fig_si:energy_fullsubs}
\end{figure}
Figure~\ref{fig_si:energy_fullsubs} shows that, even when diagonalizing the molecular Hamiltonian within the full subspace generated by the LUCJ ansatz, the resulting energy is less accurate than that obtained from the quantum computer. This implies that hardware noise, in combination with the configuration recovery procedure, generates configurations beyond those accessible to the original ansatz, which are essential for obtaining the improved energy profile.

\bibliography{abbr, references} 